\documentclass[english,preprint,aps,prb,preprintnumbers,floatfix]{revtex4}
\makeatletter

\usepackage{epsfig}
\usepackage{float}
\usepackage{float}
\usepackage{float}
\usepackage{graphicx}
\usepackage{epstopdf}
\usepackage{babel}
\usepackage{epsfig}

\makeatother

\begin{document}
\title{Correlation between structure and conductivity of stretched Nafion.}

\author{Elshad Allahyarov }

\affiliation{Department of Physics, Case Western Reserve University, Cleveland,
Ohio 44106\\
 and Joint Institute of High Temperatures, Russian Academy of Sciences
(IVTAN), Moscow, 125412 Russia}

\author{Philip L. Taylor}

\affiliation{Department of Physics, Case Western Reserve University, Cleveland,
Ohio 44106}

\begin{abstract}

We have used coarse-grained simulation methods to investigate the
effect of stretching-induced structure orientation on the proton
conductivity of Nafion-like polyelectrolyte membranes. Recent
experimental data on the morphology of ionomers describe Nafion as an
aggregation of polymeric backbone chains forming elongated objects
embedded in a continuous ionic medium. Uniaxial stretching of a recast
Nafion film causes a preferential orientation of these objects in the
direction of stretching. Our simulations of humid Nafion show that
this has a strong effect on the proton conductivity, which is enhanced
along the stretching direction, while the conductivity perpendicular
to the stretched polymer backbone is reduced. Stretching also
causes the perfluorinated side chains to orient perpendicular to the
stretching axis. 
 This in turn affects the distribution of
water at low water contents. The water forms a continuous network with
narrow bridges between small water clusters absorbed in head-group
multiplets.

\end{abstract}

\maketitle

\section{Introduction}

In their role  as proton-conducting membranes, ionomers are an
important component of many hydrogen fuel cells.  In these materials,
the interplay among the short-range interactions between the
hydrophobic backbone polymer and the hydrophilic terminal groups, and
the long-range Coulomb interactions between the electrostatic charges
on the terminal groups and the protons induces a nanophase separation
into proton-rich and proton-poor domains. A general model for the
phase morphology of ionomers has been proposed  
by Eisenberg {\it et al.}~\cite{ehm}, according to which a few head groups combine to form multiplets
 that restrict the mobility of the backbone chain segments directly
attached to them.   A spherical geometry is often assumed for these
multiplets, whose sizes are typically less than a
nanometer \cite{gohy}. The average distance between these 
multiplets is mostly dictated by the concentration of head groups
relative to that of backbone monomers. 
 These multiplets then form microdomains, and one observes a
 microphase separation that can  serve to facilitate proton diffusion.  
 Recent experimental data on the morphology of ionomers describes the
 aggregations as elongated objects embedded in a continuous ionic medium \cite{rubatat2002}.  
A goal of much ionomer research is to increase the proton
conductivity, and hence to make membranes that can operate under very low
humidity conditions, as a higher conductivity results in a higher output
power density in fuel cells containing these membranes.  

Proton conduction itself is a complex process, which strongly
depends on the thermal and mechanical history of the
membrane.  The mechanical history involves the manufacture of the membrane, which is usually achieved by one of two common procedures: solution casting or extrusion. 
The former method is used to make membranes from a solution of dissolved ionomer by allowing the solvent to evaporate from the solution. While this technique is suitable only for
 small-scale laboratory production, it has the advantage producing isotropic membranes with no residual preferential
orientation of their backbone within the plane of the membrane. 

Most commercially available ionomeric membranes are fabricated by extrusion of a 
molded sample. This leads to a preferred
orientation of the ionomer backbone \cite{barbi2003}, the extent of which 
depends on the draw speed of the extrusion. This inherent structural
anisotropy of the backbone matrix is believed to be a reason for the susceptibility to tearing
or cracking of the membrane in any swelling or drying processes. This
creates technical problems in keeping the membrane taut in the
changing temperature-humidity conditions encountered in fuel cells.  An equally important issue concerns the effect of this anisotropy on proton conductivity.  While it has generally been assumed that such effects are significant, the problem of predicting the consequences of mechanical strain remains largely unsolved.

There have been several experimental studies in which membranes were uniaxially stretched in order to probe the effects of strain on
internal morphology. Gebel {\it et al.}~\cite{gebel2005} analyzed the
form of ionic domains in unstretched and stretched membranes, and showed
that mechanical stretching induces ordering in the ionomer
backbones. Elliot {\it et al.}~\cite{elliot2000,elliot2006} detected the anisotropy in
stretched membranes by performing scattering experiments. Barbi {\it et al.}~\cite{barbi2003} used membrane elongation measurements to confirm the classical ionomer
domain model, in which inverted micelles are interconnected by channels.
As expected, uniaxial stretching of recast Nafion  causes a preferential
orientation of the Nafion backbone  in the direction of stretching, and this is
 morphologically similar to the anisotropy in extruded membranes.
 Cable {\it et al.}~\cite{cable1995}, for example,  stretched Nafion and
found the proton conductivity to be higher in the plane of the membrane
than normal to it.  A slightly different result was found by  Lin {\it
  et al.}~\cite{pinta2007}, who noted little change in transverse
conductivity on stretching, but found an improved fuel cell  
performance, as compared to Nafion 117 and unstretched recast
Nafion, as a consequence of a lowered methanol permeation rate.  
In a related experiment,  Elabd {\it et al.}~\cite{elabd}
measured the conductivities of an ionic block copolymer, and observed appreciable
anosotropy.  
 Oren {\it et al.}~\cite{oren} found that the
conductivities of membranes could be rendered anisotropic through the
alignment of suspended particles, with higher conductivity in the
alignment direction.  These studies reveal the significant impact that
organized and oriented structures can have in increasing proton
conductivity.
 While there have been several intensive simulational studies of the
 effects of  naturally occurring anistropies on the transport
 properties of Nafion-like membranes
 \cite{cui2007,blake2005,jang2004,wescott2006,vishnyakov2001}, there
 are no studies of which we are aware of the 
effects of mechanical stretching on an equilibrated sample.

 In this paper we investigate numerically to what extent induced orientation
 in the backbone polymer affects proton
 diffusion and  conductivity in Nafion-like ionomers.
  The nature of
our united-atom model for the ionomer is described in  Section II, and
the details of our simulation procedure are 
given in Section III.  Section IV reports our results for stretch-induced
morphological changes in the ionomer, while results for the
 proton conductivity are given in  Section V. In Section VI we discuss our conclusion that
stretching of  solvated membranes leads to 
 anisotropy of the conductivity, with the stretch-induced ordering of the backbone
increasing the longitudinal  protonic
 conductivity while decreasing the transverse  conductivity.

\section{Model}
 \label{sectionmodel}
Because it is necessary to model quite large volumes of material in
 order to see the types of morphological changes in which we are
 interested, it is unfortunately not feasible to perform fully
 atomistic simulations.  We thus employ the same united-atom model 
 for Nafion \cite{wescott2006,vishnyakov2001,yamamoto} as was
 used in our previous work \cite{allahyarov-1}. 
Within this approach the CF$_{2}$
and CF$_{3}$ groups of the backbone and sidechains, the ether oxygens of the sidechain, the  sulfur atom and the  O$_3$ oxygen group
of the sulfonates are modeled as Lennard-Jones (LJ) monomers of
diameter $\sigma=3.5$\AA. All the ether oxygens and fluorocarbon groups are  
assigned zero partial charges. The electrostatic charges are located
entirely on the  SO$_{3}^{-}$ sulphonate head groups and on the  H$^{+}$ protons. 

The total potential energy of the system is then given as 
\begin{equation}
U_{{\rm{total}}}=U_{\rm LJ}+U_{q}+U_{{\rm{bond}}}+U_{{\rm{angle}}}
+ U_{{\rm{dihedral}}}  .
\label{eq:1}
\end{equation}
Here the first two terms on the right-hand side are pair interactions
between nonbonded monomers. The first term is the Lennard-Jones interaction 
\begin{equation}
U_{\rm LJ}(r)=4\varepsilon_{LJ}\sum_{i>j} \left( (\sigma/r_{ij})^{12}- a
  (\sigma/r_{ij})^{6} \right)
\end{equation}
where $\varepsilon_{LJ}=0.2$ kcal/mol, and $r_{ij}= \vert  \vec r_i -
\vec r_j\vert$ is the separation distance between monomers $i$ and
$j$.   At a temperature of 300K, this value of $\varepsilon_{LJ}$ would correspond to an energy of about $k_{B}T/3$.   The parameter $a$ is 1 for hydrophobic-hydrophobic 
(HH) interactions,  and 0.5 for both  hydrophobic-hydrophilic
(HP) and  hydrophilic-hydrophilic (PP) interactions
between monomers.
In the PP case,
the LJ potential was modified to be
purely repulsive by truncating it at its minimum, where $r
=1.122 \sigma$, and raising it by the addition of an amount  $\varepsilon_{LJ}$.
These 
interaction parameters  were chosen to agree in most
instances with the Nafion model of Paddison  {\it et al.}~\cite{Paddison1}. 

The electrostatic interaction between charged sulphonate
groups and protons is
\begin{equation}
U_{q}=\sum_{i>j}\frac{q_{i}q_{j}e^{2}}{\epsilon(r_{ij}) r_{ij}}\label{eq:2}
\end{equation}
Here $q_l$ ($l=i,j$)  is +1 for
 protons, +1.1 for sulfur atoms, and $-2.1$ for the combined oxygens of the sulfonate
 groups, thus ensuring that  the total charge
  of a sulfonate head group  SO$_{3}^{-}$ is $-e$.
The distance-dependent dielectric permittivity $\epsilon(r)$ in Eq.~(\ref{eq:2}) 
reflects the fact that, at small separations, there is no intervening material between any two charges.   There is then no screening of the Coulomb interaction, and the effective permittivity is close to unity.    At modest  distances the charges will be separated by matter that may be water, ionic groups, or backbone polymer, and the energy of interaction will be reduced by the shielding effect of this material. 
 The interaction
 between sulphonates, both inside the sulphonate head
 group clusters surrounded by polymeric matrix, and between two
 different sulphonate clusters separated by polymeric matrix, is thus a
 complicated function that includes the effects of image charge distributions at the
 boundaries \cite{imagecharge_eff_diel}.  When water is present,  dielectric saturation and water immobilization effects in the hydrophilic
 phases of the ionomeric membrane \cite{kreuer2004,paul2001} add complications.   
While  the best approximation for  $\epsilon(r)$ is still a matter for debate \cite{xu}, it is
clear that the
permittivity must increase with distance. We choose the form $\epsilon(r)= 1+\epsilon_{B}((r/\sigma - 1)/(r/\sigma + 1))^{10}$, which
 smoothly increases from $\epsilon=1$ near  a chosen charge to being
the dielectric permittivity $\epsilon_{B}$ of the bulk at large distances.   Varying the value of $\epsilon_{B}$
 between 8 and 20 has only a negligible effect on all
 the observed nanophase separations in the 
 system under consideration. The effective dielectric permittivity
 $\epsilon_{\rm{eff}}$, calculated as a volume average of
 $\epsilon(r)$ over a typical multiplet domain,  appears to be in  good agreement with reported 
 dielectric constants for ionomers in Ref.~\cite{epsilon-nafion}.

The last three terms on the right side of Eq.(\ref{eq:1}) represent the
potential energy of the bonded segments of the molecules. The
two-body bond-stretching  potential
\begin{equation}
U_{{\rm{bond}}}(R)= \frac{1}{2} \sum_{\rm{all \,  bonds}} k_{b}(R-R_{0})^{2}\label{eq:3}
\end{equation}
has a simple Hookean form with stretching constant $k_{b}=700$ kcal/mol 
\AA$^{2}$ (roughly 1200 $k_BT$/\AA$^{2}$ at room temperature) 
 and unstretched bond length $R_{0}$=1.54 \AA \, (0.44$\sigma$). 
For the three-body angle-bending potential,   
\begin{equation}
U_{\rm{angle}}(\theta)=\frac{1}{2}\sum_{\rm{all \, bond \, pairs}}
k_{\theta}(\theta-\theta_{0})^{2},\label{eq:4} 
\end{equation}
we use an equilibrium
bending angle $\theta_{0}=110^{\circ}$ and  a bending force constant $k_{\theta}=120$ kcal/mol
deg$^{2}$ (or about 200  $k_BT$/deg$^{2}$ ). 
Finally, the dihedral (four-body) component of the total energy is written as
\begin{equation}
U_{\rm{dihedral}}(\phi)=\frac{1}{2}\sum_{\rm{all \, bond \, triplets}}
k_{\phi}\left(1-d \cos(3\phi)\right),
\label{eq:5}
\end{equation}
with the parameters having values  $d=-1$ $(+1)$ and $k_{\phi}$=18.1 kcal/mol
$\sim$ 10.8 $k_BT$ (6.2  kcal/mol $\sim$ 3.7$k_BT$) for backbone
(sidechain) segments, respectively.

 We  consider two models for the ionomer.  In the first model, which
 is a highly-simplified representation of a completely dry membrane,
 each proton is permanently attached to a sulfonate group.   Each
 side-chain  head-group is thus 
 electrostatically neutral, and consists of two dipoles, connected
 head to tail.  We refer to this as the bound-proton model.

In the second, and more realistic, model, the proton is more loosely
bound to its sulfonate counter-ion by a Coulomb attraction.  We refer
to this as the solvated model. 
The activation energy needed to overcome the 
electrostatic attraction between the oppositely charged H$^+$ and SO$_3^-$
ions, and thus detach the proton, depends on the
average distance between sulfonates and on the water content $\lambda$ in
the membrane. This parameter $\lambda$ is defined as the number of water
molecules per
 sulfonate head group, and varies between 0 and 25 for solvated
 membranes.  It is
 believed that  $\lambda \ge 5$ is  sufficient to loosen the bonds tying the protons to the head groups and to achieve full dissociation.  These free protons contribute to the
 conductivity principally through diffusion and through the Grotthus mechanism. When $\lambda < 5$, hopping diffusion, in which
 protons make a transition from one sulfonate to an adjacent one, becomes the main contributor to the conductivity.      

 We use a simple point charge (SPC) fluid
 \cite{jang2004,water_parameters} to model the water molecules in our
simulations.   These move in a medium having a distance-dependent dielectric permittivity
 $\epsilon(r)$ that we assume to be independent of $\lambda$.   This includes the case of $\lambda=0$, where in the absence of SPC water molecules
 the proton conductivity occurs
 only through the hopping mechanism.

\section{Simulation details}

The simulations were performed in stages. The first step was to grow
a continuous backbone with attached pendant sidechains by using Monte-Carlo 
techniques. The bond length, and the bending and dihedral angles for this
initial polymer were fixed to be  $R_{0}$,  $\theta_{0}$ and  $\phi_{0}$
respectively. After molecular-dynamics (MD) runs of a few picoseconds, which were used to equilibrate the system, as this usually
needed to respond to the strong steric repulsion of overlapping monomers,
the constraints on angles and bonds were removed.
Following this,  the side chains were detached from
the backbone skeleton in an approach previously adopted by Vishnyakov
\cite{rivin-vishnyakov}, and the backbone skeleton was cut into
segments \cite{glotzer2002}. This greatly increases the relaxation 
rate of the ionomer, and allows a rapid equilibration of the system. 
The system  then consists of
$N_s=1000$ sidechain segments 
and $N_b=1400$ backbone segments.  Taking into account that the sulfonic
acid groups are hydrophilic while the ethers and fluorocarbon groups
are hydrophobic \cite{Paddison1}, we adopted the following 
coarse-grained representation. The side chain architecture is
written as 7$H$+3$P$ for a dry membrane and 7$H$+2$P$ for a humidified
membrane. Here the letters $H$ and $P$ stand for hydrophobic and
hydrophilic monomers correspondingly.
The backbone segments are fully hydrophobic with a 14$H$ architecture.

Molecular-dynamics runs with a Langevin thermostat were performed for the segmented polymer
for time
periods up to 50ps in a constant-$NVT$ ensemble.  We imposed periodic
boundary conditions for a cube of side $L=30\sigma$ and used the
Lekner summation method \cite{Lekner} to handle the long-range electrostatic
interactions between charged particles. In order to verify that the system
had not become trapped  into a metastable glassy state, we repeated each run with
several different initial configurations.

In the next stage of the simulations, the segments were reassembled
 back into the  branched chain characterizing the original Nafion. 
This was achieved by the simultaneous  introduction of  the bonds and
 angular constraints  described by  Eqs.~(\ref{eq:3})--(\ref{eq:5})
 between the ends of each backbone segment, 
which united them into a single chain.  Similar bond and angular
 constraints between the tail monomer of each detached side chain and the median 
monomer of every backbone segment connected the side chains to the
 backbone polymer. To avoid the formation of star-like branched
polymers,  only a single occupancy of the backbone attachment sites was permitted.   The unequal numbers of backbone segments and side chain segments resulted in the formation of a polymer in which there were varying numbers of backbone monomers between the points of attachment of the side chains.  
The simulation was then resumed, and run until a new equilibrium was achieved.
Once the system was fully equilibrated, the statistically averaged
 quantities of interest were gathered during the next 2--3ns of run time for the initial case of unstretched membranes.

The stretched samples were constructed by the application of a set of forces designed to emulate the effects of a longitudinal stress.    
 The 
continuous backbone polymer was considered as a set of $N_b$ connected segments, and a weak force ${\bf f_\pm}$ was applied to the end monomer of each segment.  The direction of this force was in the positive $z$ direction for the end with the greater value of its $z$ coordinate, and in the negative  $z$ direction for the end with the lesser value of its $z$ coordinate in an extended-zone scheme where the periodic boundary conditions were not applied.  This force 
tends to align the backbone segments in the $z$ direction without having significant influence on the side-chain dynamics.     
The strength of ${\bf f_\pm}$  was restricted to values much
smaller than the forces arising from the bond, angle, and dihedral
potentials.

 It usually took 50ps to equilibrate the stretched membrane.   Following this, runs of 2--3ns were performed to gather statistically significant results for the morphology of the stretched material.   For studies of the proton conductivity, a weak electric field was applied to the equilibrated stretched membrane, and
additional 50ps runs were performed to reach steady-state conditions before 
Non-Equilibrium MD  (NEMD) simulation runs of 2--3ns were undertaken.  In this way the effects of both stretching and applied fields could be determined.

\section{Results for Stretched-Membrane Morphology}

\begin{figure}
\includegraphics*[width=0.54\textwidth]{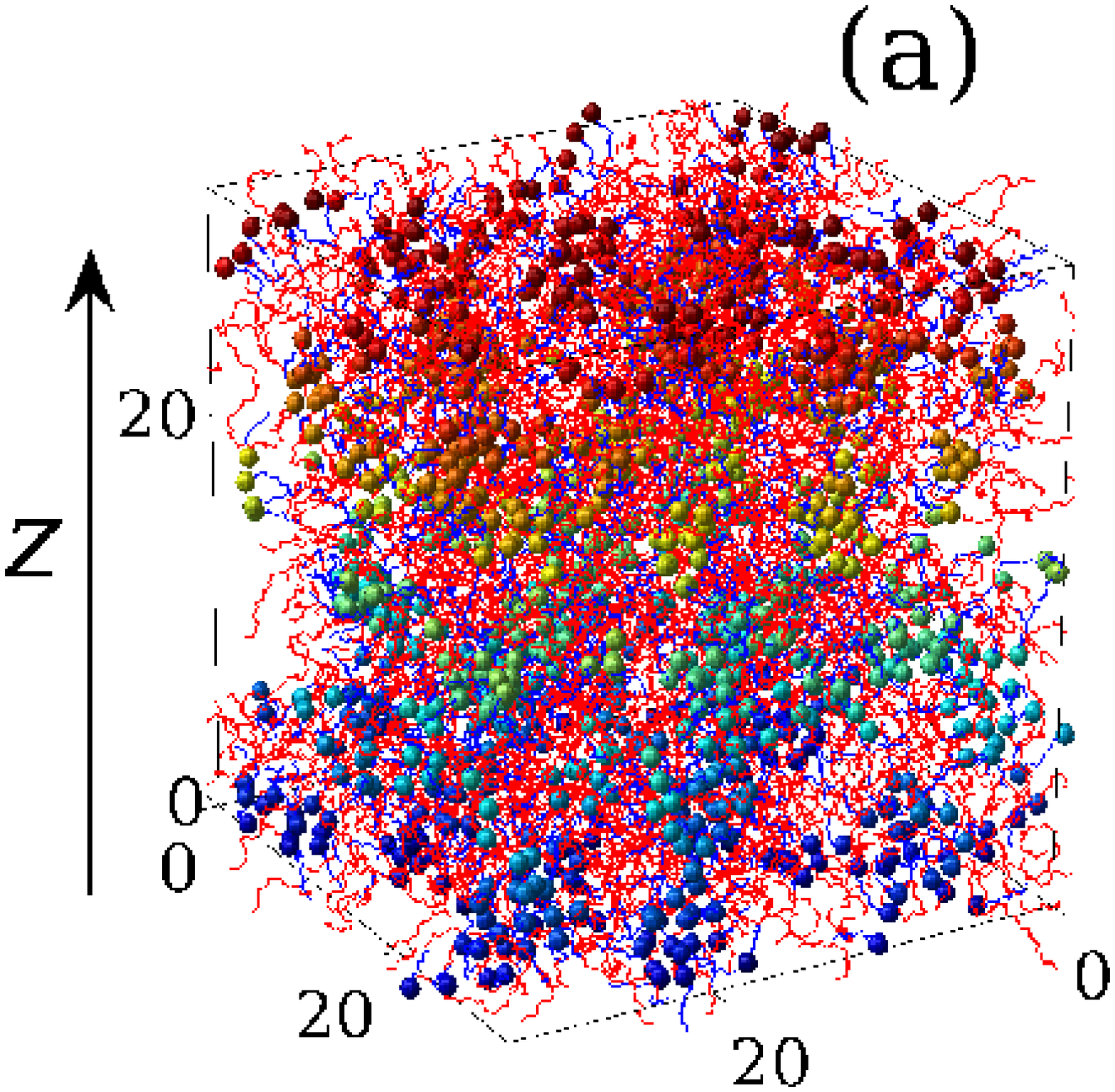}
\includegraphics*[width=0.54\textwidth]{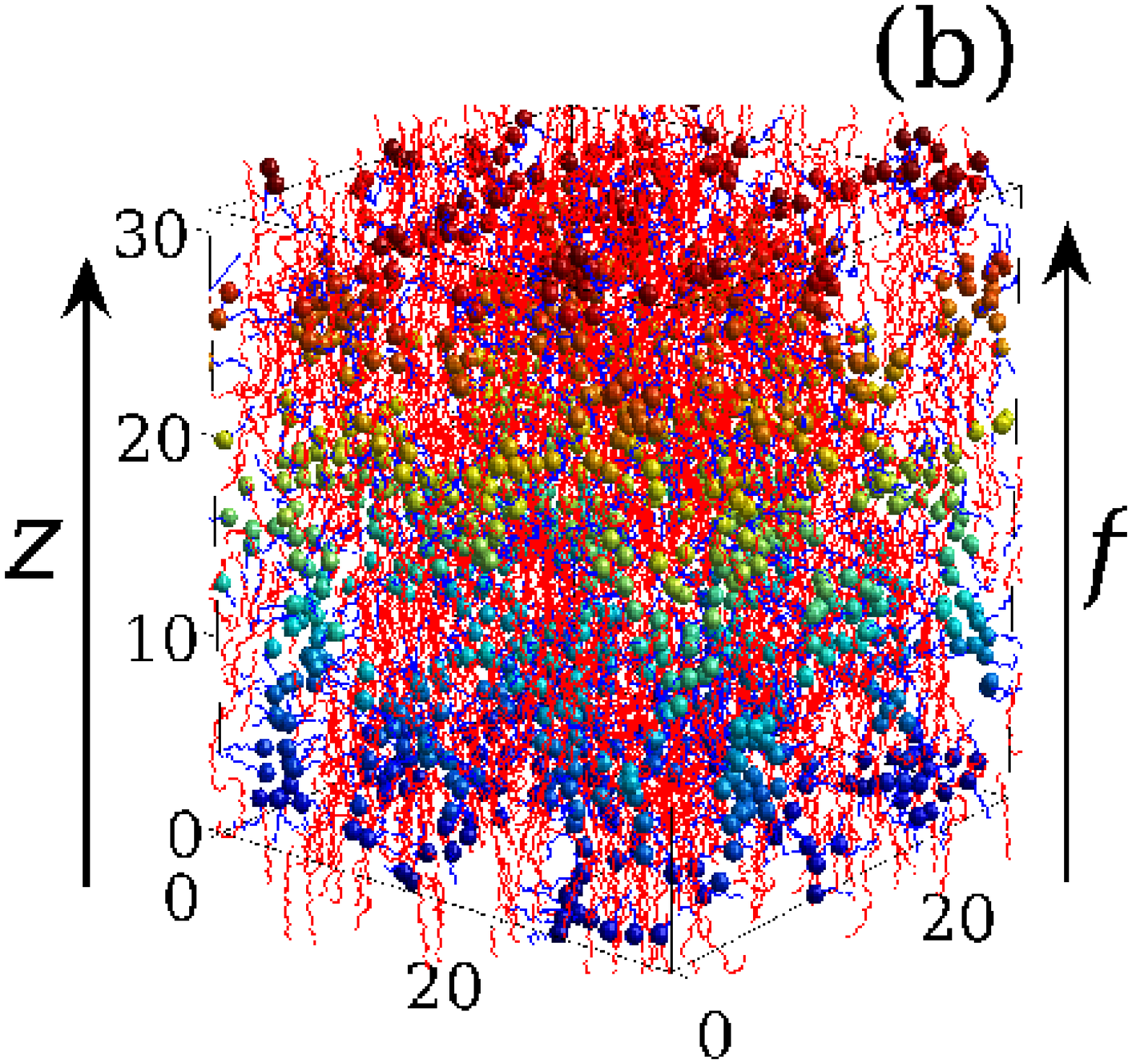}
\caption{(Color online) A typical snapshot  of
  the simulation box for (a) unstretched ($f$=0)  and (b) stretched ($f$=1) 
  Nafion membrane with water content $\lambda=3$.  The  stretching force $f$ in (b)  is directed
  along the $z$ axis.  Colored beads  represent the end-group oxygens of side
  chains. Pendant side chains  and neutral backbone polymer are
drawn by lines (blue and red, respectively, in online
  version). Different bead colors correspond 
to different bead altitudes, with blue for  beads
at the bottom of the simulation box and red for
beads at the top of the box. The size of all structural
elements is schematic rather than space filling. 
\label{fig-1}}
\end{figure}

\begin{figure}
\includegraphics*[width=1.\textwidth]{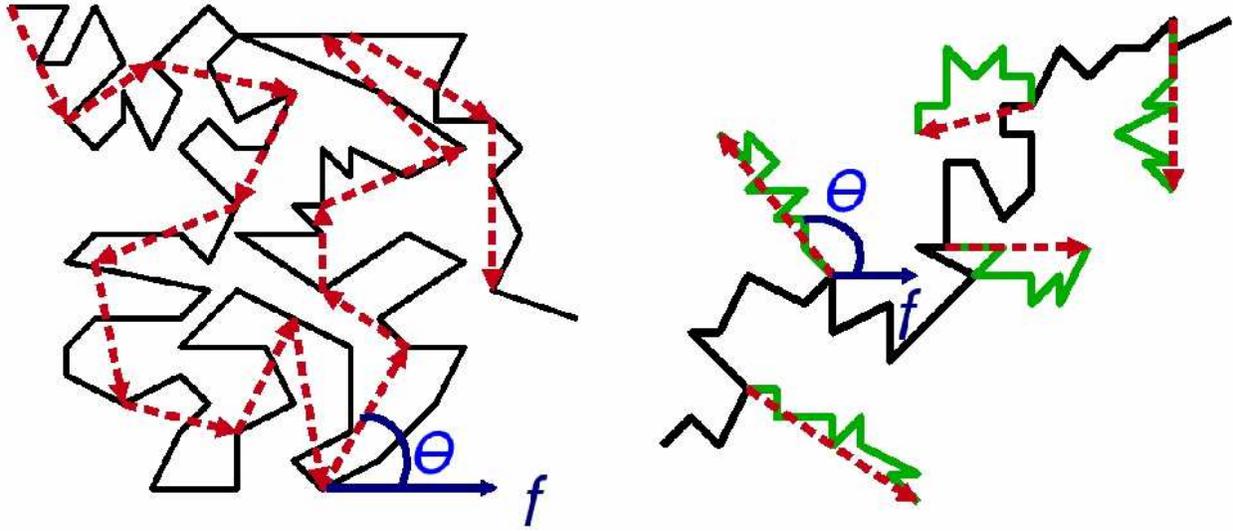}
\caption{(Color online)  This schematic illustration explains the
  calculation of order parameters $S_{b}$ 
  for backbones (left picture), and $S_{s}$ for sidechains
  (right picture).  The continuous black line represents the backbone polymer of the ionomer. Shorter
  gray lines (green in online version) represent the
  sidechains. The dashed lines are vectors constructed to show the orientation of backbone
  segments and sidechains, with $\theta$ being the angle  between the orientation vector and stretching direction.    
   \label{fig-2} }
\end{figure}

A snapshot of an equilibrated membrane with a water content of $\lambda=3$ molecules per head group
and in the absence of an applied stretching force is
given in Fig.~\ref{fig-1}(a). The hydrophobic backbone and the side chains are shown as lines (red and blue respectively in the online version).  The spheres
indicate the sulfonate head groups of the sidechains, and show 
cluster-like aggregations  of different sizes and 
shapes, in  agreement with experimental observations \cite{mauritz2004}. 
A moderately stretched membrane with  dimensionless stretching force $f$=1 is shown in
Fig.~\ref{fig-1}(b). The polymer backbone shows
ordering in the direction of the applied force, which is parallel to the  $z$ axis. We
determine the induced backbone ordering $S_b$ and sidechain ordering
$S_s$ by calculating the order parameter 
\begin{equation}
 S_j= \frac{1}{2} \left( 3 \left< \frac{1}{N_b} \sum_{i=1}^{N_b} cos^2\theta_i \right > -1
\right),
\label{order}
\end{equation}
an approach commonly used in liquid crystal systems.  
Here $j=b,s$ for the backbone polymer  and sidechains respectively,
the angular brackets
 $\left < ... \right >$ denoting  statistical averaging over the $NVT$
ensemble.    The orientational angle $\theta$ for each segment  is defined as the angle  
 between the vector connecting the first and last monomers of the 
segment and the applied force ${\bf f}$, as illustrated in  Fig.~\ref{fig-2}.
 The backbone matrix is  considered as a set of $N_b$ segments of
 14 monomers each. If $\theta$ is larger than $\pi/2$, the 
order of monomer numbering along the segment is reversed  for a proper
 accounting its orientation. 
In a similar manner we define the orientational angle  $\theta$ for sidechains as also shown in Fig.~\ref{fig-2}. Here $\theta$ is the angle between the
stretching force ${\bf f}$ and the vector pointing from the first monomer of the side
chain towards its terminal group. 

\begin{figure}
\includegraphics*[width=0.7\textwidth]{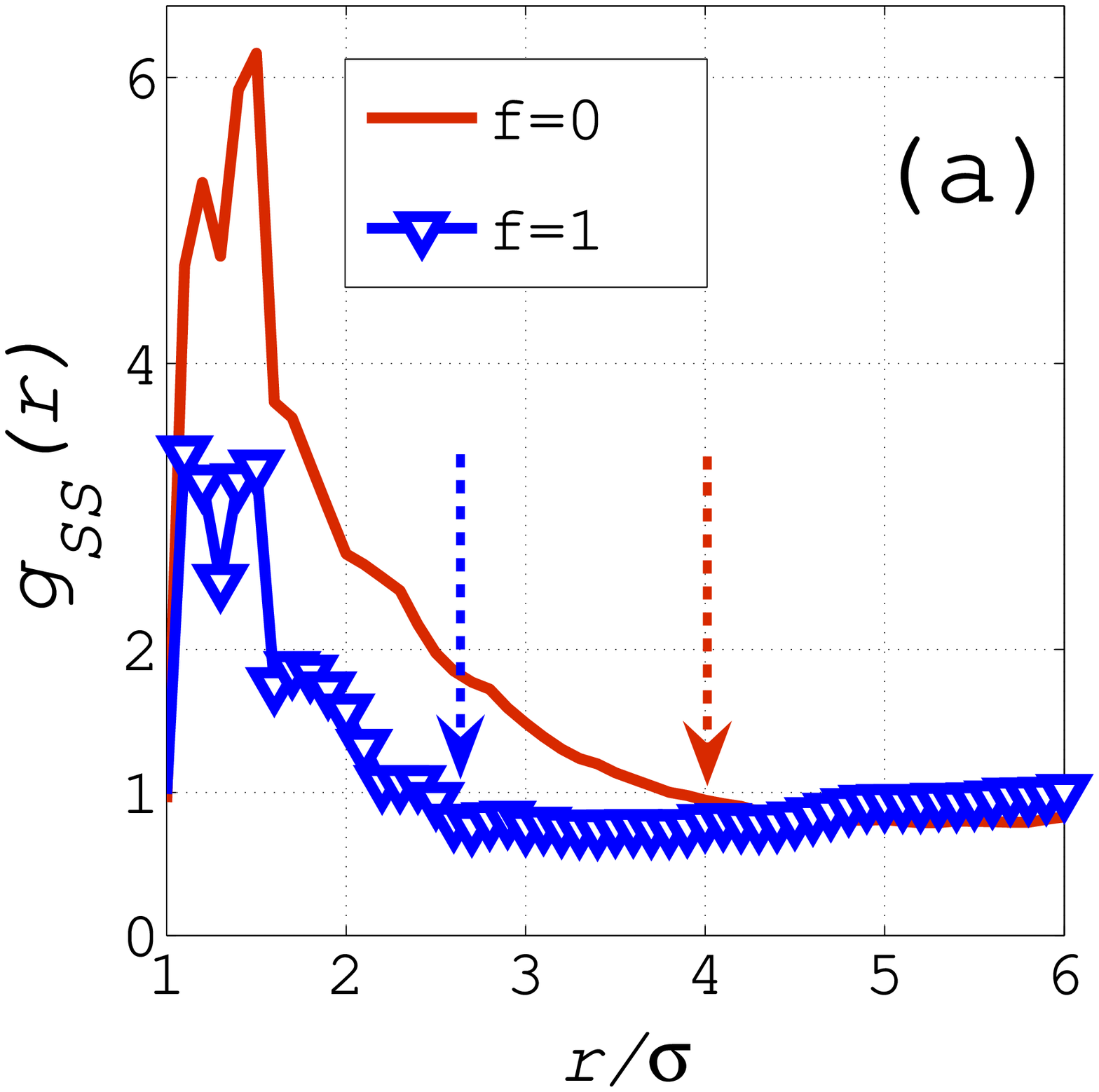}
\includegraphics*[width=0.7\textwidth]{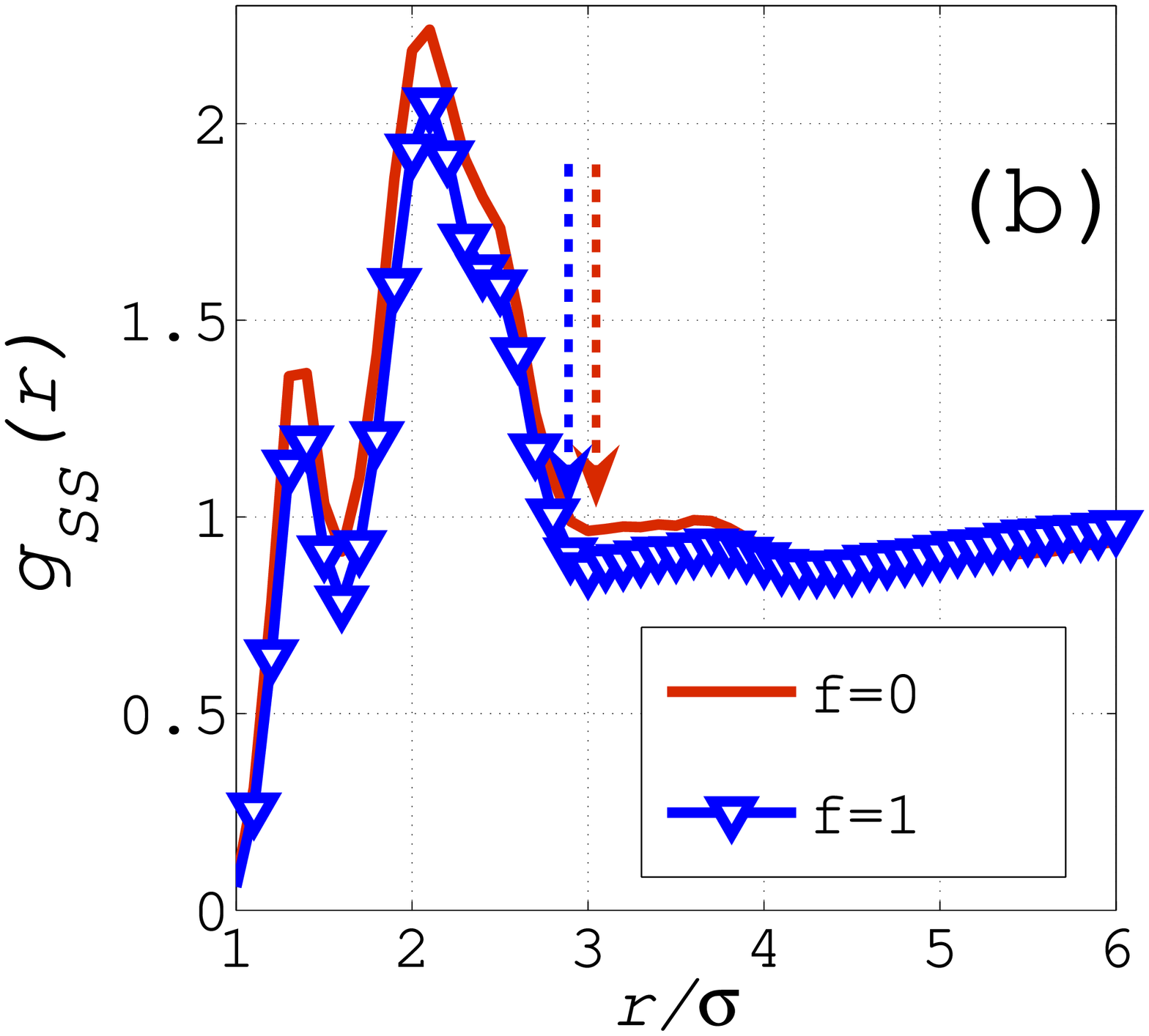}
\caption{(Color online)  This figure shows the effect of strain on the sulfur-sulfur pair
  distribution function $g_{ss}(r)$  for (a) the bound-proton model and (b) the solvated ionomer with $\lambda=5$. Dashed arrows show the average size of multiplets.
   \label{fig-5}}
\end{figure}

 The order parameters $S_b$ and $S_s$, calculated  from  Eq.~(\ref{order}), are  
shown in Fig.~\ref{fig-3}.    
As  expected, membrane stretching leads to ordering in the ionomer
conformation. The backbone ordering, shown in Fig.~\ref{fig-3}(a),   is greatest in the bound-proton model and decreases in the solvated case as the water content $\lambda$ increases for a fixed force ${\bf f}$.  Figure~\ref{fig-3}(b) reveals the completely different response of the side chains
to the stretching force. When the bound-proton membrane is stretched,
the sidechains tend to orient themselves parallel to the stretching
force. However when a solvated membrane is stretched, the sidechains are
more likely to be perpendicular to the stretching force, in 
accord with previous experimental results \cite{chourdak}. 
 A possible explanation for these different responses to stress lies in the effect of the humidity on multiplet formation.    In the absence of water, the dipolar interactions between sulfonates leads to the formation of dense multiplets containing 10--20 terminal groups, as seen in
Fig.~\ref{fig-5}(a). The Coulomb forces of attraction between these dipoles 
 limit the amount  of molecular relaxation of the sidechains in response to
stretching.  The backbone deformation is thus followed by the rearrangement of
multiplets via elongation in one direction and shrinking in the other
direction. This is evident from a comparison of the curves
for stretched and unstretched membranes in Fig.~\ref{fig-5}a.  

 In contrast, when the membrane is solvated, the water content lowers the densities of the multiplets.  These then coalesce to form larger
 clusters in which a mixture of solvent molecules and conducting
 protons gives  the head groups more dynamic flexibility. As a
 result, the sidechains can achieve a gain in entropy by rearranging themselves
 perpendicular to the direction of the applied stress.
This is the reason why  only a slight deformation of the multiplet
 geometry for the solvated membrane is seen in Fig.~\ref{fig-5}(b). 
In summary, the order parameter of the membrane 
    depends on its hydration level, the absorbed solvent being an obstacle for the
    backbone ordering $S_b$, but facilitating the side-chain
 ordering $S_s$.   

\begin{figure}
\includegraphics*[width=1\textwidth]{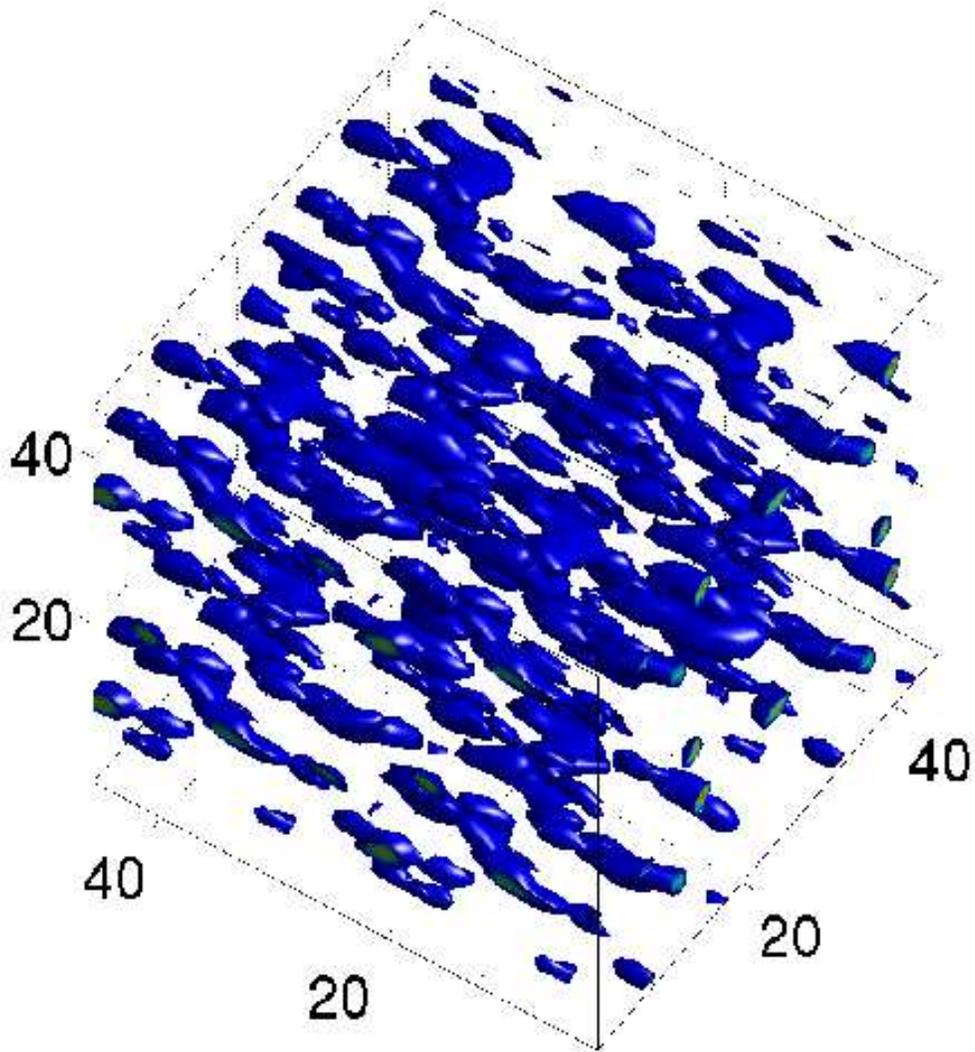}
\caption{(Color online) Instantaneous distribution of water clusters in a
  stretched, solvated  membrane for stretching force $f=1$ and water content $\lambda=5$ water molecules per sulfonate.
\label{fig-6}}
\end{figure}

The most important aspect of the effect of stress on morphology concerns the formation of channels through which proton transport might occur.   An indication of this is shown in 
Fig.~\ref{fig-6}, in which the water distribution in a stretched ionomer is
displayed.   The water molecules congregate in elongated clusters in association with the distribution of side-chain head-groups, and are oriented in the direction of the applied stress.
  During the course of the simulation,
the clusters create temporary bridging connections allowing the water to diffuse easily
back and forth along the stretching direction. This morphology
and dynamics of the water undoubtedly facilitates proton diffusion parallel to
the stretching direction.

\begin{figure}
\includegraphics*[width=1.\textwidth]{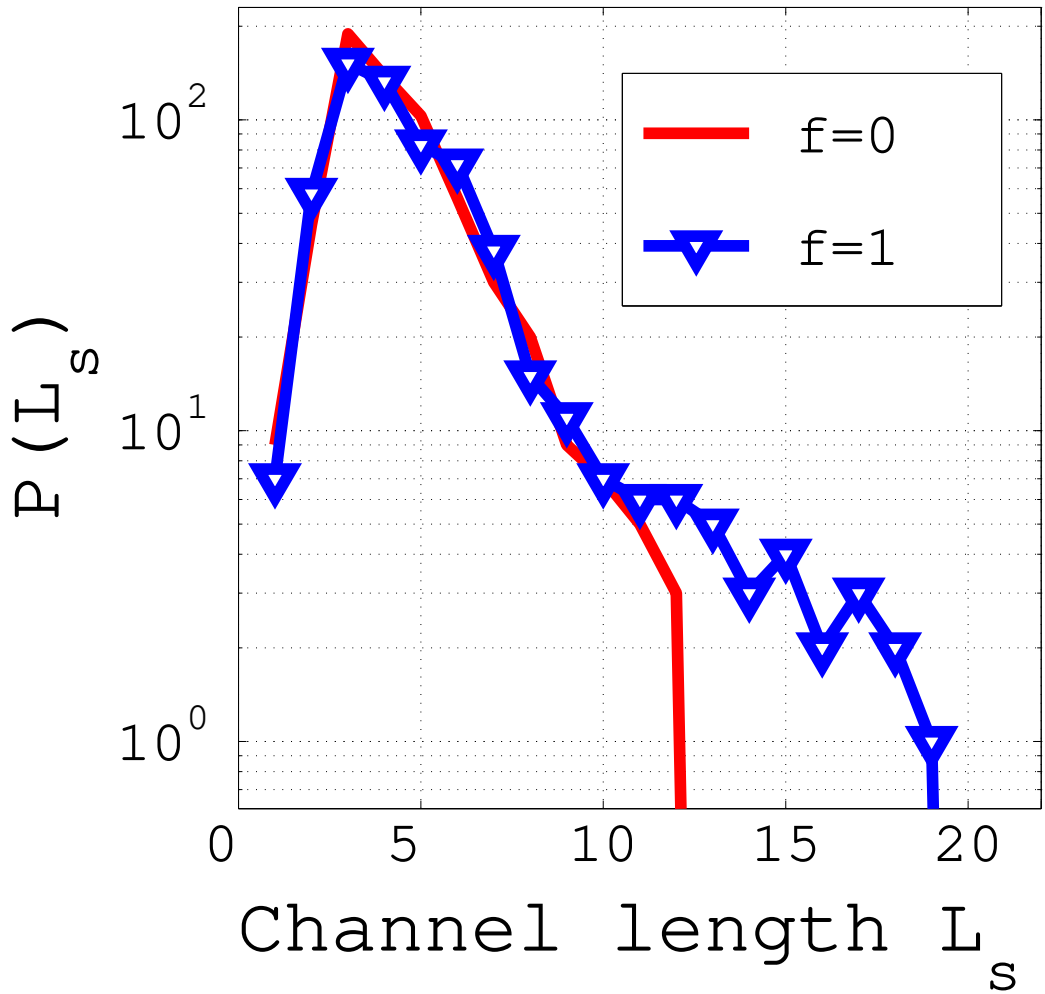}
\caption{(Color online) The distribution function $P(L_s)$  of sulfonate channel lengths $L_s$ 
   for unstretched ($f$=0) and stretched ($f=1$) solvated  membranes having a
  water content $\lambda=5$.
  \label{fig-8}}  
\end{figure}

In order for proton conduction to occur, it is necessary for there to be a continuous network of pathways through the hydrophilic material in a membrane.  This makes it useful to have a quantitative measure of the connectivity of the hydrophilic channels.
To this end we have developed a  method
 \cite{our-efield} aimed at detecting hydrophilic 
pathways in a network of hydrophilic clusters. The only input
parameter for this method is the maximum
separation distance parameter $r_m$ between neighboring particles 
in the pathway. Our method calculates the distribution function
$P(L_S)$ of the hydrophilic channel length  $L_S$, which is defined as the 
distance between the initial  and end points of the pathway. 
We applied this method to determine how the stretched-membrane morphology alters the
channeling in the hydrophilic network of clusters.  
The distribution function $P(L_S)$ when
$r_m$ is chosen to be $2\sigma$ (which is 7\AA, the average separation between
sulfonates in Nafion) is shown in Fig.~\ref{fig-8}, both for stretched and
unstretched membranes.  Both curves have a similar short-range behavior
and show a maximum at a most probable channel length equal to about 3$\sigma$.  This
is roughly the average size of the 
sulfonate multiplets in solvated membranes, as was seen in Fig.~\ref{fig-5}(b). The long-range   
behavior of $P(L_S)$, however, shows a dependence on the  mechanical treatment of the membrane. In a stretched membrane  the
hydrophilic-sulfonic channels in the membrane are 50\% longer
than in an unstretched membrane. We note that the calculated  $P(L_s)$
 is an instantaneous  measure of the pathways in a network of clusters.   Because of the constantly changing patterns in the distribution, which are caused by
 the diffusion of ions and water and the motion of head groups, the real pathways are effectively
 longer, and can even traverse the whole system. 
Consequently one expects that the protons in a stretched membrane can continuously
travel along the stretching axis $z$. We 
explore the resulting protonic conductivities in the next section.

\section{Results for the Conductivity of a Stretched Membrane}

In this section we analyze the effect of membrane stretching on the
protonic current. The electrical conductivity $\chi$ due to proton current in a  solvated membrane can
be obtained by performing NEMD simulations to find the current $\bf j$ in the presence of an external
electrostatic field $\bf E$ through the relation 
\begin{equation}
{\bf j}= \chi({\bf E})\, \bf E.
\label{eq2}
\end{equation}
Here the induced  current density $\bf j$  is given by  
\begin{equation}
j = (1/V) \displaystyle\sum_{i=1}^{N_S} q_i v_{i,\bf E}
\label{eq3}
\end{equation}
 where $v_{i,\bf E}$ is the E-field component of the velocity of the $i^{\rm th}$ ion, and
 $N_S$ is the total number of protons in the system. 

\begin{figure}
\includegraphics*[width=1.0\textwidth]{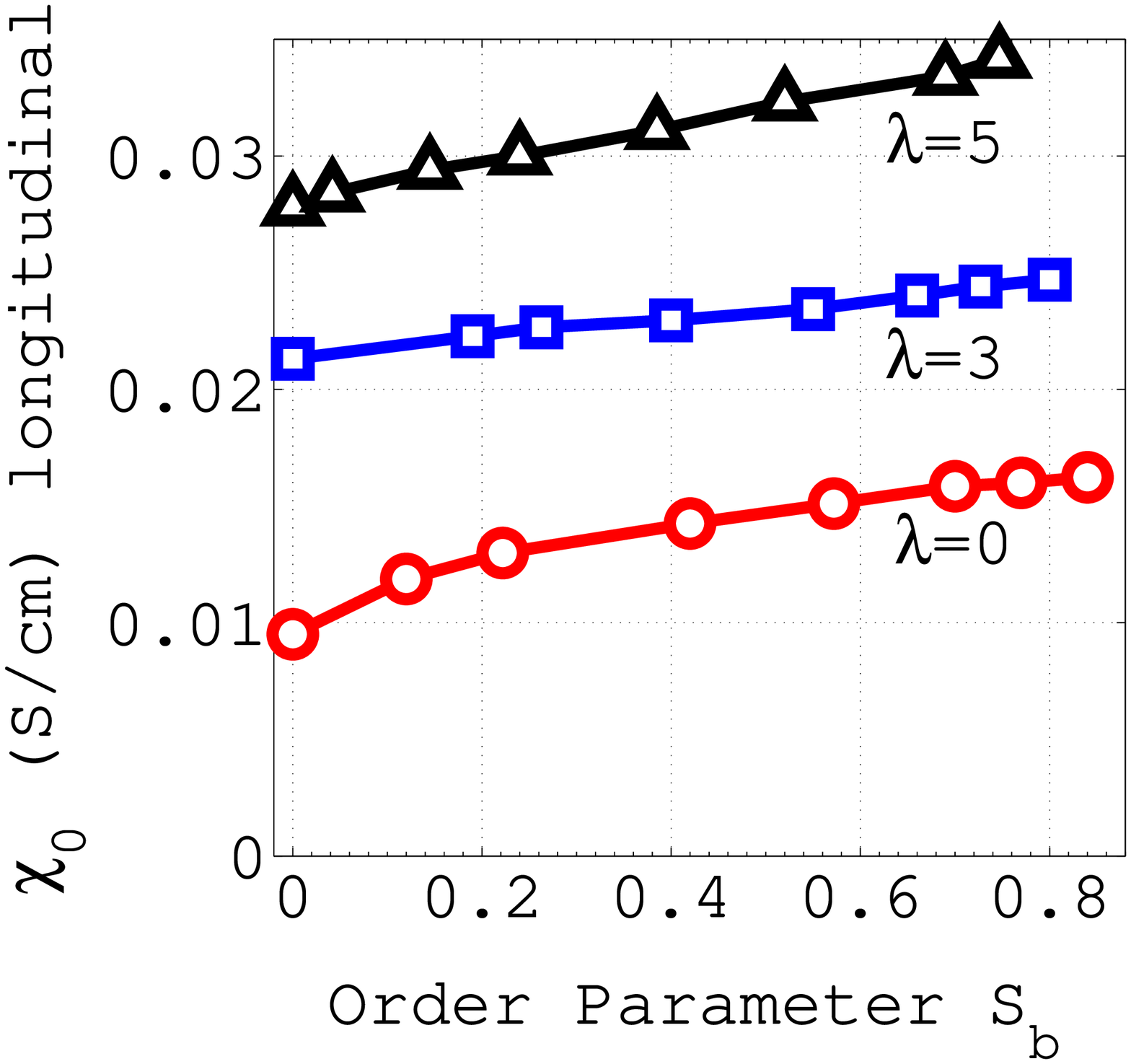}
\caption{(Color online) Protonic conductivity $\chi_0$ of a stretched membrane
  along the stretching direction as a function of the backbone order parameter
  $S_b$ for different membrane solvations. The water-content parameters
  $\lambda$ are indicated below the corresponding lines.  
\label{fig-9}} 
\end{figure}

\begin{figure}
\includegraphics*[width=1.0\textwidth]{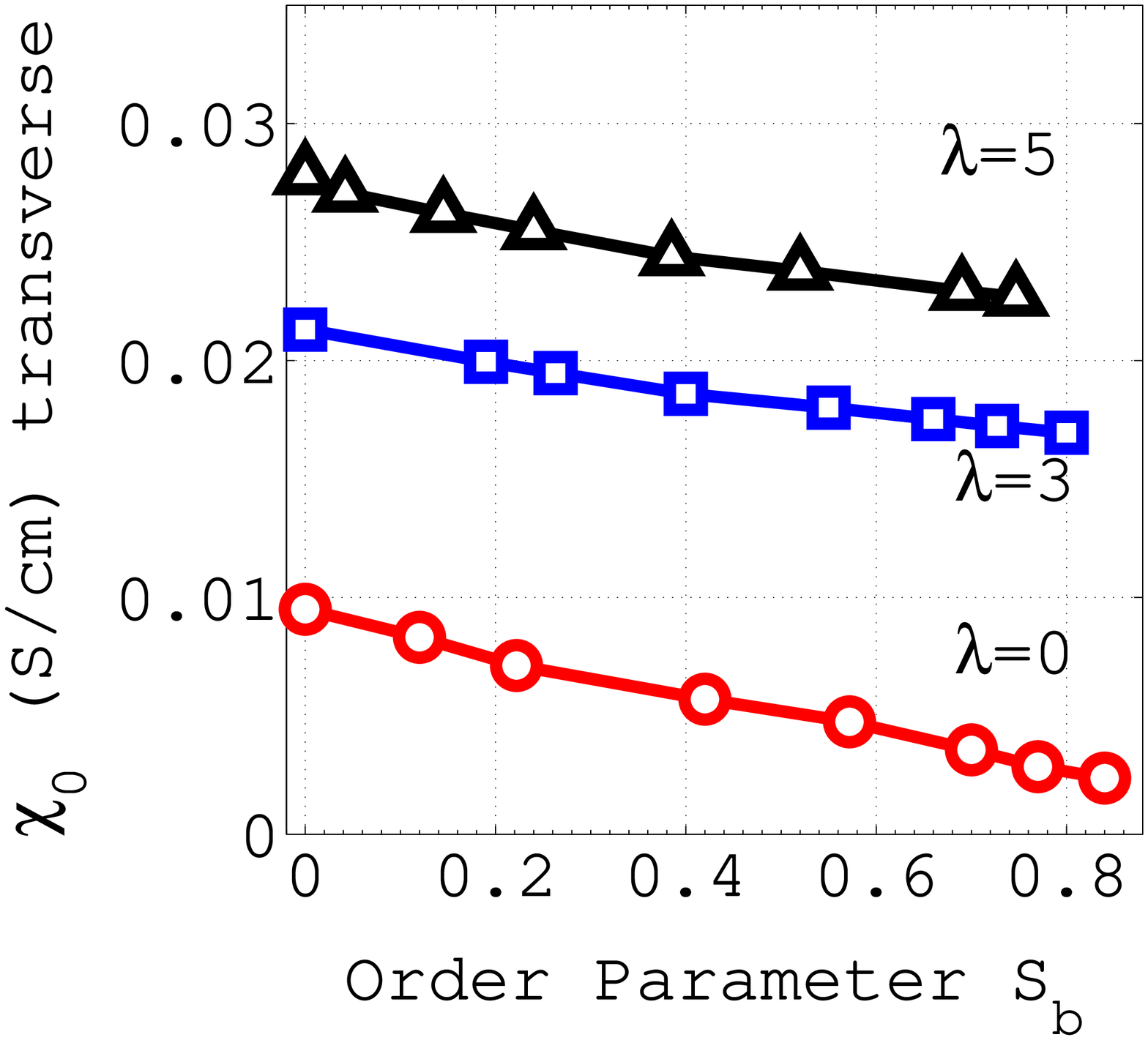}
\caption{(Color online)Protonic conductivity $\chi_0$ of a stretched membrane
  perpendicular to the stretching direction as a function of the backbone order parameter
  $S_b$ for different membrane solvations. The water-content parameters
  $\lambda$ are indicated below the corresponding lines.  
\label{fig-10}} 
\end{figure}

Because a strong applied electric field may itself have an effect on the morphology, we isolate the effects of stretching by examining only the limit $\chi_0$ of $\chi({\bf E})$ as ${\bf E}\to 0$.  The longitudinal and transverse conductivities were evaluated by applying   fields  parallel and  perpendicular  to the stretching direction,
and extrapolating the resultant conductivities to find their values at zero applied field.  The calculated values of $\chi_0$ for the case where ${\bf E}$ is along the membrane stretching axis  are shown in Fig.~\ref{fig-9} as a function of the
membrane order parameter $S_b$ for different
solvation parameters $\lambda$.  It is seen that ordering in the membrane
backbone increases the conductivity, with a dependence of  $\chi_0(S_b)$
on $S_b$ that  is close to being   linear for all water contents $\lambda$.    
Conversely, when we examine the transverse protonic conductivity $\chi_0$ across the stretched  membrane, which is
shown in  Fig.~\ref{fig-10}, we find that the conductivity decreases with increasing
backbone ordering. These two trends can be understood as a direct
consequence of the channel-like nanophase separation along the
stretching direction. While the protons can freely travel along these channels as a consequence of the
water cluster bridgings, there are no continuous pathways for them to travel
across the stretching direction.   This anisotropy in the conductivity is most pronounced in the dryer samples.

\section{Discussion and Conclusion}
The simulation results presented in the previous sections show that the
uniaxial stretching of a Nafion-like ionomer should be expected to have a significant effect 
on its proton conductivity. The conductivity along the  stretching
direction is considerably higher than its value in an unstretched membrane, while the
proton diffusion perpendicular to the  stretching axis is strongly
reduced.  When a water-containing sample is stretched,  the perfluorinated side chains exhibit a
tendency to be oriented perpendicular to the stretching force direction.
The sulphonate clusters deform as the stretching
 is increased, and show a long-range
ordering in their distribution across the membrane. 
This  affects the distribution of absorbed water, which forms  a
continuous  network with narrow bridges. 
     
\begin{figure}
\includegraphics*[width=0.7\textwidth]{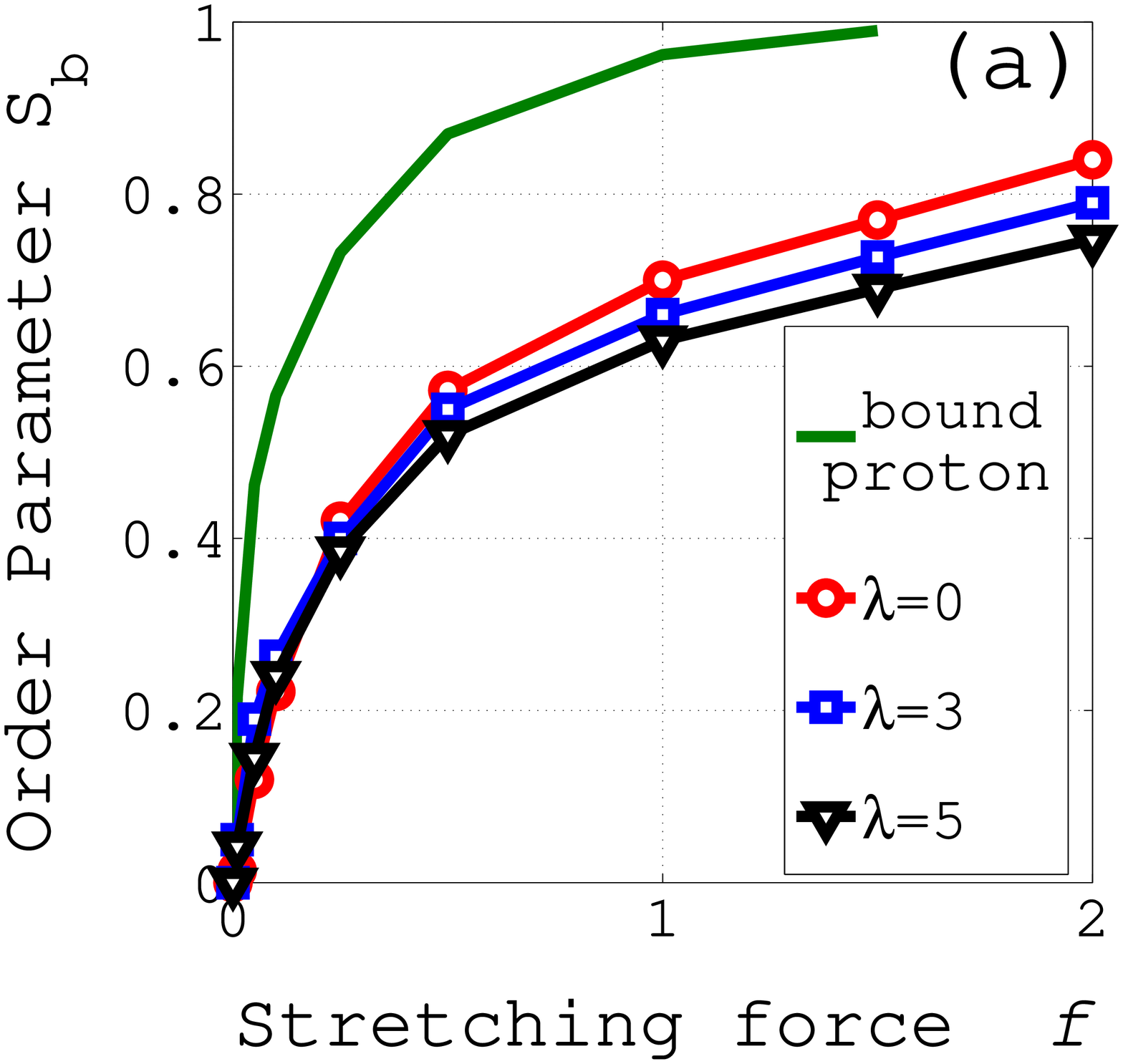}
\includegraphics*[width=0.7\textwidth]{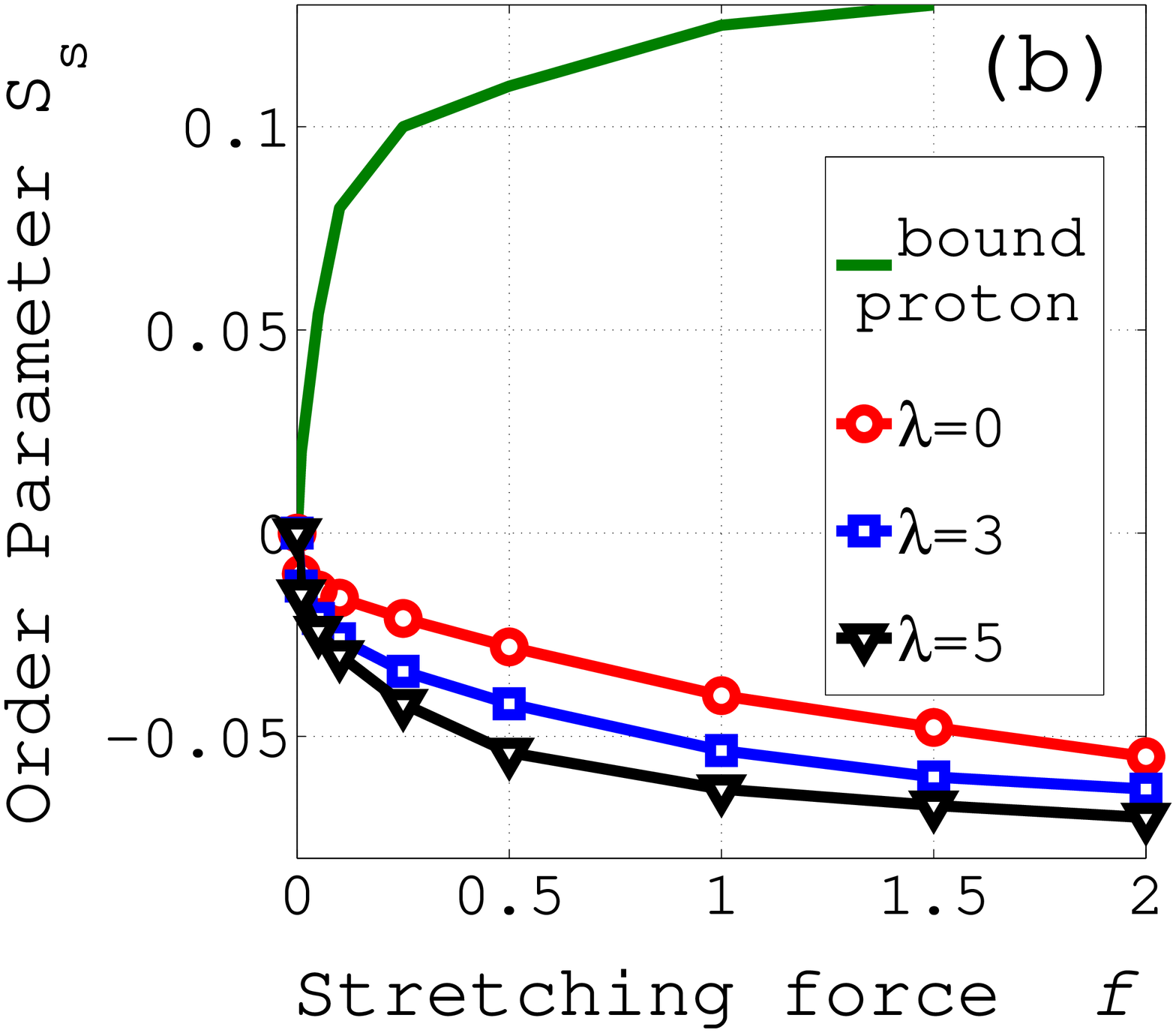}
\caption{(Color online) Backbone order parameter $S_b$ (a) and
  sidechain order parameter $S_s$ (b) as functions of
  applied stretching force $f$ for the bound-proton model and for different water contents
  $\lambda$. 
  \label{fig-3} }
\end{figure}

 The effect  of the water is different for the backbone and sidechain
ordering parameters. As seen from Fig.~\ref{fig-3}, water is an
obstacle to backbone ordering $S_b$ parallel to the direction of elongation. However, water enhances the sidechain ordering $S_s$
perpendicular to the stretching direction.  
For a given stretching force ${\bf f}$, the order parameter $S_b$ of the membrane  backbone
depends on the solvation parameter $\lambda$.

The simulations reported here were carried out at room temperature, and under 
conditions of constant stress, and we made no attempt  to investigate how the
anisotropy in the protonic conductivity depends on the temperature at
which the membrane was stretched.    Experimental measurements of crystallinity, however, indicate that
membrane  stretching at
elevated temperatures can lead to different results from ambient-temperature stretching \cite{trevino, cable1995}.
It appears that uniaxial extension of
 membranes at temperatures above the boiling temperature of water results 
 in a material with an oriented morphology that persists after the
 samples are removed from the extensional stress. 
Also,  higher-temperature compression-molded  ionomers exhibit a backbone
crystallization when stretched \cite{bagrodia}.  It would thus be of interest to pursue further studies to enlarge on those presented here.

Finally, we note that with regard to the issue of improving fuel-cell performance by using stretched membranes, our results indicate that the observed increased proton conductivity occurs only in the direction of stretching.  As a practical matter, the desired direction of proton transport is across the membrane, and so the production of a membrane with enhanced conductivity due to stress may be difficult  to arrange.

\begin{acknowledgments}
We acknowledge stimulating discussions with P. Pintauro and R. Wycisk,
who introduced us to their membrane stretching
experiments.  E. A. thanks M. Litt for valuable comments on equilibrium ionomer morphologies.
This work was supported by the US Department of Energy under grant
DE-FG02-05ER46244, and was made possible by use of facilities at the
Case ITS High Performance Computing Cluster and the Ohio Supercomputing
Center.
\end{acknowledgments}

\end{document}